\documentclass{aa}  

\bibpunct{(}{)}{;}{a}{}{,} 

\usepackage{graphicx}
\usepackage{txfonts}
\usepackage{pdflscape}
\usepackage[utf8]{inputenc}
\usepackage{xcolor}
\usepackage{comment}
\usepackage{hyperref}
\usepackage{arydshln}
\usepackage{dblfloatfix}
\usepackage{cancel}
\usepackage{multirow}

\newcommand{\asas}{ASAS-SN}
\newcommand{\gaia}{{\it Gaia}}
\newcommand{\sptz}{{\it Spitzer}}
\newcommand{\event}{AT2021uey}
\newcommand{\gaiaevent}{Gaia21dnc}

\newcommand{\pan}{Pan-STARRS}
\newcommand{\ztf}{ZTF}
\newcommand{\lco}{LCO}
\newcommand{\pylm}{{\it pyLIMA}}
\newcommand{\mlm}{{\it MulensModel}}
\newcommand{\bes}{{\it Besan\c{c}on}}

\newcommand{\kojima}{TCP J05074264+2447555}
\newcommand{\dkv}{Gaia22dkv}

\newcommand{\ohp}{OHP/Mistral}
\newcommand{\pepsi}{LBT/PEPSI}

\newcommand{\pylimass}{{\it pyLIMASS}}
\newcommand{\mist}{{\it MIST}}

\begin{document} 

\title{\event{}: A planetary microlensing event outside the Galactic bulge}

\author{
Ban, M.,\inst{35}
Voloshyn, P.\inst{2,3}, 
Adomavi\u{c}ien\.{e}, R.\inst{4},
Bachelet, E.\inst{6}, 
Bozza, V.\inst{7,8},
Brincat, S. M.\inst{9},
Bruni, I.\inst{10}, 
Burgaz, U.\inst{11},
Carrasco, J. M.\inst{12,28,34}, 
Cassan, A.\inst{5},
\u{C}epas, V.\inst{4},
Cusano, F.\inst{10}
Dennefeld, M.\inst{5},
Dominik, M.\inst{13},
Dubois, F.\inst{14},
Figuera Jaimes, R.\inst{15,33},
Fukui, A.\inst{16,17},
Galdies,~C.\inst{18,19},
Garofalo, A.\inst{10},
Hundertmark, M.\inst{20},
Ilyin, I.\inst{32}
Kruszy\'{n}ska, K.\inst{1,26}, 
Kulijanishvili, V.\inst{21},
Kvernadze, T.\inst{21},
Logie,~L.\inst{14},
Maskoli\={u}nas, M.\inst{4},
Miko{\l}ajczyk, P. J.\inst{1,22}, 
Mr\'{o}z, P.\inst{1}, 
Narita, N.\inst{16,17,23},
Pak\u{s}tien\.{e}, E.\inst{4},
Peloton, J.\inst{3}, 
Poleski,~R.\inst{1}, 
Qvam, J. K. T.\inst{24},
Rau, S.\inst{14},
Rota, P.\inst{7,8}
Rybicki, K. A.\inst{1,25}, 
Street, R. A.\inst{26},
Tsapras, Y.\inst{20},
Vanaverbeke,~S.\inst{14},
Wambsganss,~J.\inst{20},
Wyrzykowski, \L{}.\inst{1,29},
Zdanavi\u{c}ius, J.\inst{4},
\.{Z}ejmo, M.,\inst{30},
Zieli\'{n}ski, P.\inst{27}, and
Zola, S.\inst{31}
}

\institute{
University of Warsaw, Astronomical Observatory, Warszawa, Poland \and
Taras Shevchenko National University of Kyiv, Ukraine \and
Université Paris-Saclay, CNRS/IN2P3, IJCLab, 91405 Orsay, France \and 
Institute of Theoretical Physics and Astronomy, Vilnius University, Vilnius, Lithuania \and
Institut d'Astrophysique de Paris (IAP), Sorbonne Universit\'e, CNRS, Paris, France \and
IPAC, Mail Code 100-22, Caltech, 1200 E. California Blvd., Pasadena, CA 91125, USA \and
Dipartimento di Fisica "E.R. Caianiello", Universit{\`a} di Salerno, Fisciano, Italy \and
Istituto Nazionale di Fisica Nucleare, Sezione di Napoli, Napoli, Italy \and
Flarestar Observatory, San Gwann, Malta \and
INAF, Osservatorio di Astrofisica e Scienza dello Spazio di Bologna, Bologna, Italy \and
School of Physics, Trinity College Dublin, Dublin, Ireland \and 
Institut de Ci\`encies del Cosmos (ICCUB), Universitat de Barcelona (UB), Mart\`{\i} i Franqu\`es 1, E-08028 Barcelona, Spain \and 
University of St Andrews, Centre for Exoplanet Science, SUPA School of Physics \& Astronomy, UK \and
Public observatory Astrolab IRIS, Zillebeke, Belgium \and
Millennium Institute of Astrophysics MAS, Nuncio Monsenor Sotero Sanz 100, Of. 104, Providencia, Santiago, Chile \and
Komaba Institute for Science, The University of Tokyo, Tokyo, Japan \and
Instituto de Astrof\'{i}sica de Canarias (IAC), Tenerife, Spain \and
Institute of Earth System, University of Malta, Malta \and
Znith Astronomy Observatory, Malta \and
Zentrum f{\"u}r Astronomie der Universit{\"a}t Heidelberg, Astronomisches Rechen-Institut, Heidelberg, Germany \and
Evgeni Kharadze Georgian National Astrophysical Observatory, Abastumani, Georgia \and
Astronomical Institute, University of Wroc{\l}aw, ul. M. Kopernika 11, 51-622, Wroc{\l}aw, Poland \and
Astrobiology Center, Tokyo, Japan \and
Horten videreg\r{a}ende skole Horten, Norway \and
Department of Particle Physics and Astrophysics, Weizmann Institute of Science, Rehovot, Israel \and
Las Cumbres Observatory Global Telescope Network, 6740 Cortona Drive, Suite 102, Goleta, CA 93117, USA \and
Institute of Astronomy, Faculty of Physics, Astronomy and Informatics, Nicolaus Copernicus University in Toru\'{n}, Toru\'{n}, Poland \and
Institut d'Estudis Espacials de Catalunya (IEEC), Esteve Terradas, 1, Edifici RDIT, Campus PMT-UPC, 08860 Castelldefels (Barcelona), Spain \and
National Centre for Nuclear Research, Pasteura 7, PL-02-093 Warsaw, Poland \and
Janusz Gil Institute of Astronomy, University of Zielona Gora, Lubuska 2, 65-265 Zielona Gora, Poland \and
Astronomical Observatory, Jagiellonian University, Orla 171, 30-244 Krak\'{o}w, Poland \and
Leibniz-Institut f\"{u}r Astrophysik Potsdam (AIP), An der Sternwarte 16, 14482 Potsdam, Germany \and
Instituto de Astrof\'isica, Facultad de F\'isica, Pontificia Universidad Cat\'olica de Chile, Av. Vicu\~na Mackenna 4860, 7820436 Macul, Santiago, Chile \and
Departament de Física Qu\`antica i Astrof\`isica (FQA), Universitat de Barcelona (UB), Mart\`i i Franqu\`es 1, E-08028 Barcelona, Spain \and
Independent researcher, Warszawa, Poland
}

\titlerunning{\event{}}
\authorrunning{Ban, Voloshyn, et al.}

\date{}

\abstract{We report the analysis of a planetary microlensing event \event{}. The event was observed outside the Galactic bulge and was alerted by both space- (\gaia{}) and ground-based (\ztf{} and \asas{}) surveys. From the observed data, we find that the lens system is located at a distance of $\sim$1 kpc and comprises an M-dwarf host star of about half a solar mass, orbited by a Jupiter-like planet beyond the snowline. The source star could be a metal-poor giant located in the halo according to the spectral analyses and modelling. Hence, \event{} is a unique example of the binary-lens event outside the bulge that is offered by a disc-halo lens-source combination.}
\maketitle

\section{Introduction} \label{sec:introduction}
Observations of gravitational microlensing events have the potential to find exoplanets and the event rate is by far the highest in the direction of the Galactic bulge \citep{Paczynski1991}. Hence, the microlensing surveys were focusing their observing efforts on the Galactic bulge (because of the highest event rate) and the Magellanic Clouds \citep[in order to verify if low-mass black holes can be a main component of dark matter;][]{Paczynski1986} until a few years ago, almost all microlensing planets were found toward the Galactic bulge. The exception is a planet \kojima{}~Lb \citep[also called Kojima-1~Lb;][]{Nucita2018,Fukui2019} and a planet \dkv{}~Lb, which is the first microlensing planet discovered by \gaia{} \citep{Wu2024}. Here we present a discovery of a third microlensing planet outside the Galactic bulge, called \event{}~Lb. The event was found independently by the \gaia{} satellite \citep{Prusti2016} and two ground-based surveys: the Zwicky Transient Facility \citep[\ztf{};][]{Bellm2019,Masci2019} and the All Sky Automated Survey for SuperNovae \citep[\asas{};][]{Shappee2014}. The planetary anomaly is covered by ZTF and ASAS-SN observations in a survey mode. Hence, this planet can be used in a future study to derive how planet frequency and properties of planetary systems change with the Galactic position. \dkv{}~Lb is one of the statistical sources for the planet frequency with the Galactic position, but \cite{Wu2024} indicates the alternative possibility of M-dwarf planetary system other than the planetary system with a Sun-like host as they tentatively concluded. In contrast, Kojima-1 has no data from photometric surveys over the planetary anomaly, there are only two epochs from the ASAS-SN that are just after the anomaly \citep{Nucita2018}. Therefore, Kojima-1~Lb cannot be used for a statistical study of the Galactic distribution of microlensing planets (at least using standard methods).

We note that the Galactic latitude of \event{} ($(l, b)=(77.^{\circ}91$, $-19.^{\circ}06)$) is twice larger than for Kojima-1 ($(l, b)=(178.^{\circ}76, -9.^{\circ}33)$) and \dkv{} ($(l, b)=(287.^{\circ}37, -8.^{\circ}41)$). The doubled latitude drastically reduces the density of stars along the line of sight, so the event occurrence rate also decreases. The baseline magnitude of the \event{} source is fainter than the sources of Kojima-1 and \dkv{}, and the photometric anomaly is observed before the main peak of the light curve whilst Kojima-1 and \dkv{} show the anomalies around and after the main peak, respectively. The event with a fainter source and the anomaly before the main peak is a more severe condition to be alerted for the follow-up observation because of the relative photometric noise and the risk of late alerting to cover the anomaly. Among the microlensing exoplanets archived in NASA Exoplanet Archive\footnote{\url{https://exoplanetarchive.ipac.caltech.edu/}}, only $\sim$16\% of them is the case of successfully observed anomaly before the main peak. The average source brightness in $I$-band of those events is $\sim$1-mag brighter than the other cases (i.e. anomaly appears around the main peak or latter). \event{} hits such a lower occurrence rate in both the target region and the anomaly coverage, and the relatively fainter source and the less anomaly coverage make the light curve more challenging to analyse.

It is well known that planet statistics \citep{Winter2020} change as a function of host star metallicity \citep{Fischer2005}, mass \citep{Eggenberger2007}, and multiplicity \citep{Johnson2010}. Much less is known about how planet properties change with host star population or position in the Galaxy even though $\sim$5,000 exoplanets are known\footnote{Retrieved from NASA Exoplanet Archive and Encyclopaedia of exoplanetary systems (https://exoplanet.eu/) cross-matching as of February 2025.}. Most of these planets were found by the Kepler satellite in its original field \citep{Christiansen2022} which allows studying planet frequency as a function of position along one direction only. Kepler planets prevent a full understanding of architecture of planetary systems because all these planets were found using a single technique -- transits in this case. Each planet detection technique has its own intrinsic biases and limitations and transits efficiently detect only close-in planets and the sensitivity increases with increasing planet-to-star radius ratio. On the other hand, the microlensing technique \citep{Mao1991,Gould1992b} is discovering only wide orbit planets and its sensitivity increases with increasing planet-to-star mass ratio.

All exoplanets found via the microlensing technique except for \kojima{}~Lb, \dkv{}~Lb and \event{}~Lb are located either in the Galactic bulge or Galactic disc because the event occurrence is attributed to the stellar population density \citep{Kiraga1994}. The location of the microlensing events allows studying if the planets are more common in the bulge or in the disc toward the bulge. Unfortunately, it is hard to derive if a particular microlensing planet is in the bulge or a few kpc away and in the disc. The distances to lenses can be found routinely if one compares the microlensing event lightcurve seen from several different observatories that are separated by a significant fraction of 1~AU \citep{Gould1992a,Yee2015a}; hence, the combination of the ground-based and space-based observatories is the primary choice. The \sptz{} satellite \citep{Zhu2017} has been carrying out microlensing observations for a few years. The goal of these observations was to derive bulge-to-disc planet frequency ratio. However, \sptz{} is not perfectly suited for such an experiment due to its small field of view. The small field view necessitates a complex scheduling system to avoid bias in observations and ensure optimal resource utilisation \citep{Yee2015b}. The studies of the Galactic distribution of microlensing planets were also presented by \citet{Penny2016} and \citet{Koshimoto2021} but they based the distance estimates on parallaxes measured from the ground-based data. The final analysis of the whole \sptz{} microlensing dataset has not yet been published and the ground-based estimates of planet frequency as a function of Galactic position are of low accuracy.

Recently, the large-scale photometric surveys gained the ability to find microlensing events outside the bulge. These surveys are: the fourth phase of the Optical Gravitational Lensing project \citep[OGLE-IV;][]{Udalski2015,Mroz2020}, \ztf{} \citep{Rodriguez2022}, \asas{}, \gaia{} \citep{Wyrzykowski2023}, and VVV \citep{Husseiniova2021}. The event rate per object is getting smaller as we move further away from the Galactic plane and the event rate per sky area is falling even more sharply \citep{Mroz2020}. Only a small fraction of microlensing events show clear planetary signatures and combination of Galactic latitude of $-19.^{\circ}06$ with a planetary signal make \event{} a very rare event.

In this paper, we report the discovery of the planetary microlensing event \event{}. In \S\ref{sec:observation}, we explain the background of the event observations, and then the light curve is analysed in \S\ref{sec:parameters}. In \S\ref{sec:results}, we first describe that we find several possible scenarios for source distance and luminosity class. In order to better constrain source properties (and hence lens properties), we perform a simulation that is based on a Galactic model. Finally, the analysis of the lens properties are summarised in \S\ref{sec:conclutions}.

\begin{table}
\centering
\caption{\event{} alert and baseline object properties from the \gaia{} DR3 catalogue \citep{Brown2021}.}
\label{tab:source}
\begin{tabular}{|l|l|}
	\hline
	Parameters & Values \\ \hline\hline
	Alert & 2021-07-27 20:51:30 (GMT) \\
	$(RA, Dec)_{J2000}$ & 21:38:10.81, +26:27:59.65\\
	$(l, b)$ & $77.^{\circ}91$, $-19.^{\circ}06$ \\ \hdashline
	Baseline $G$-magnitude & $15.429\pm0.003$ \\
	Parallax [mas] & $0.438\pm0.047$ \\
	$\mu_{(RA,Dec)}$ [mas/yr] & $(-7.912, -5.027)\pm(0.045, 0.029)$ \\
	RUWE$^*$ & $1.478$ \\
	Distance [kpc] & $2.99\pm0.08$ \\
	\hline
\end{tabular}\\
\raggedright
\vspace{6pt}
* RUWE : Renormalised Unit Weight Error which indicates \\
\hspace{0.62in}the plausibility of parallax estimation.\\
\hspace{0.62in}It ideally distributes around $1.0$, and $>1.4$ is\\
\hspace{0.62in}regarded as no signal detection or a problematic\\
\hspace{0.62in}astrometric solution \citep{Lindegren2021}.
\end{table}

\section{Observations} \label{sec:observation}
The \ztf{} survey \citep{Bellm2019, Masci2019} announced the first alert at the position of \event{} on 11 June 2021. The alert was named ZTF18abktckv, and the confirmation of the microlensing nature of the alert was made three months later by the Fink broker\footnote{\url{https://fink-broker.org}} \citep{fink} due to the peculiar shape of the signal. Thanks to Fink's classification of this event as a microlensing candidate, the event AT2021uey caught the attention of the authors of this paper. The \asas{} survey \citep{Shappee2014} alerted the event/transient on 7 July 2021. The alert was named ASASSN-21mc. The \asas{} alert corresponds to the rapid increase of the observed light curve at the anomaly part whilst the \ztf{} alert was offered much earlier by sensing a gradual increase of the observed light curve at the early stage of the event. These alerts were offered by the good observation timing and sensitivity, and they did not become a direct trigger of the follow-up observations.

The \gaia{} Science Alerts system \citep{Hodgkin2013, Hodgkin2021} alerted the same event on 27 July 2021 as \gaiaevent{}\footnote{\url{http://gsaweb.ast.cam.ac.uk/alerts/alert/Gaia21dnc/}}. We summarise the general information about the \gaia{} alert and the \gaia{} DR3 \citep{Prusti2016, Hodgkin2021} information about the catalogue star at the same position in Table~\ref{tab:source}. The alert was recognized to be a candidate microlensing event, which triggered follow-up observations by smaller telescopes. However, the alert and the follow-up observation ended up observing only the main peak since the alert was issued after the anomaly peak. Follow-up data have been uploaded and calibrated using the Black Hole Target Observation Manager (BHTOM\footnote{\url{https://bhtom.space/}}). BHTOM is a tool for coordinated observations and processing of photometric time-series based on the open-source TOM Toolkit by LCO \citep{Street2024} and uses the Cambridge Photometric Calibration Server (CPCS, \citet{Zielinski2019, Zielinski2020}) and CCDPhot (Miko\'{l}ajczyk et al., in prep.) that are the main kernels for data processing and calibration. The PSF photometry it computes is standardised to APASS or SDSS catalogues. The information about the telescopes involved in the photometric follow-up observations is provided in Table~\ref{tab:observers01}.

The imaging observations of the \event{} field started after the \gaia{} alert. The Observing Microlensing Events of the Galaxy Automatically Key Project (OMEGA\footnote{\url{https://lco.global/science/keyprojects/}}) started automatic observations the 11 August 2021 via the Microlensing Observing Platform (MOP\footnote{\url{https://mop.lco.global}}). Hundred of SDSS-$g'$ and SDSS-$i'$ images have been collected. We also used two low-resolution spectrograph: \ohp{}\footnote{\url{https://ohp.osupytheas.fr/telescope-de-193cm/}} and a high-resolution spectrograph \pepsi{}\footnote{\url{https://pepsi.aip.de/}}.

\begin{figure*}
\centering
\includegraphics[width=\textwidth]{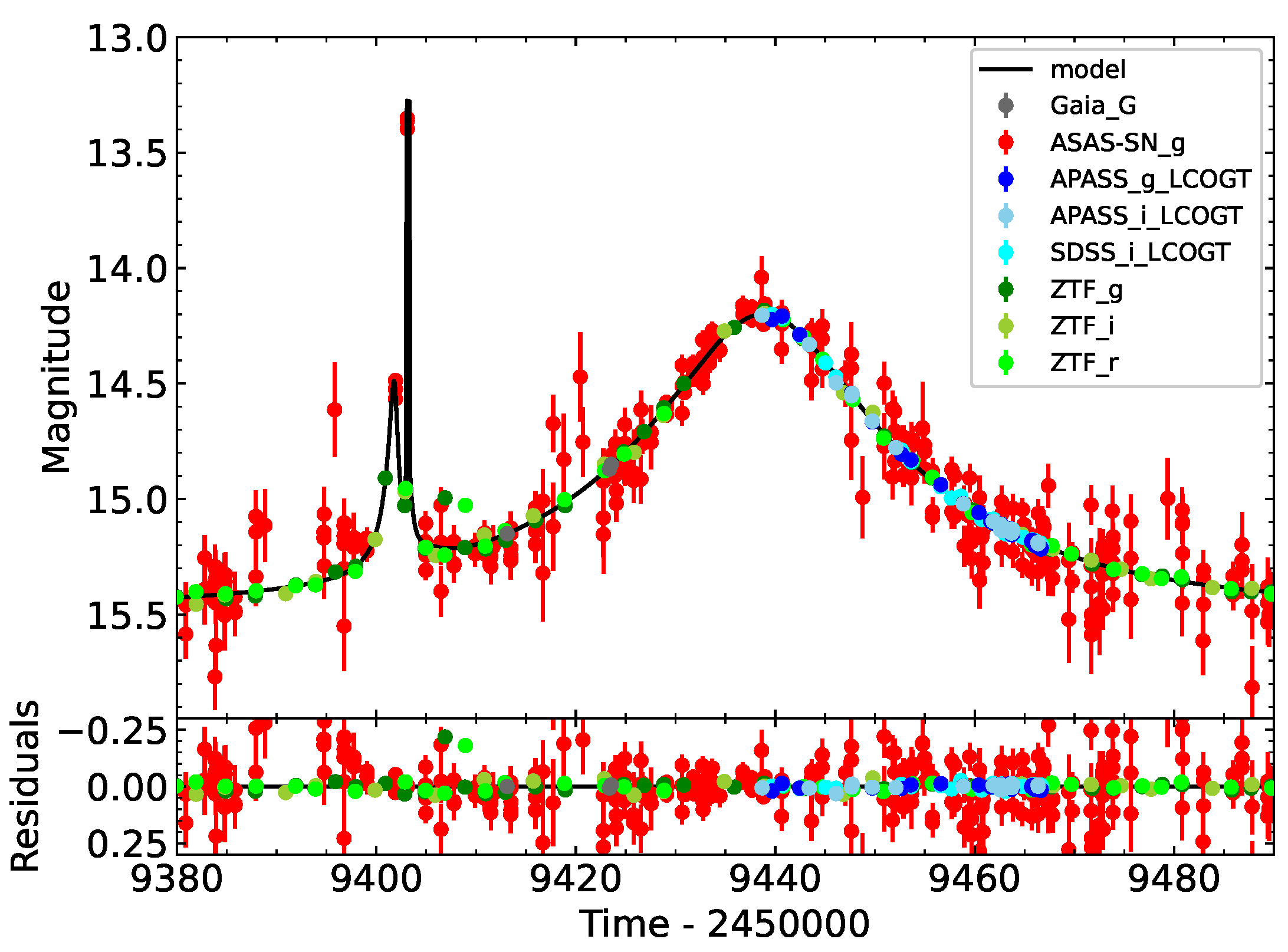}\\
\caption{Photometric data and fitted light curve of the event \event{}. The binary lens event is assumed. We show a zoomed-in light curve for the anomaly in Figure \ref{fig:anomaly}.}
\label{fig:ssbl}
\end{figure*}

The light curve of \event{} is presented in Figure~\ref{fig:ssbl}. The main peak of the event is well covered by different datasets and is 1.5~mag brighter than the baseline. There is a 3-day long anomaly at $\mathrm{HJD}=2459400$ (i.e., 4 July 2021, 40 days before the main peak -- see Figure~\ref{fig:anomaly}). During the anomaly, photometry was obtained only by the \asas{} and \ztf{} surveys. In total, nine epochs were collected during the anomaly. Among those anomaly epochs, the \asas{} survey collected two epochs that are 0.9~mag brighter than the main peak. We check if the anomaly could be produced by a nearby star or an instrumental effect. First, we search for nearby stars in the \gaia{} DR3 and \pan{} \citep{Chambers2016} catalogues. The nearest object is at a separation of $6.4$~arcsec and is too faint to be detected by either \asas{} or \ztf{} surveys. The nearest star detectable by these surveys is 49.0~arcsec away from the event centroid and has $G = 16.18~\mathrm{mag}$. We check that this star is not variable in either \asas{} or \ztf{} data. Second, we inspect the images for the anomaly epochs and have not seen any instrumental effects that could affect the photometry. Third, we note that there are two \ztf{} anomaly epochs taken in different filters and they show similar brightness shift relative to the baseline, which is one more argument for microlensing origin of the anomaly. In summary, we do not see any evidence against microlensing origin of the anomaly.

Photometric data have been cleaned up to remove outliers. We use the leave-one-out cross validation method with $\chi^2_i<9$ limitation for each epoch $i$. This cleaning is done except for the anomaly part and assuming a single-lens-single-source model. Finally, we decide to use photometric data from \gaia{}, \asas{}, \ztf{}, and \lco{}. Other datasets ensure that there are no other anomalies, but they do not constrain event parameters significantly.

\begin{table*}
\footnotesize
\centering
\caption{Photometric observations of the event \event{}}
\label{tab:observers01}
\begin{tabular}{l l l l l}
\hline
Facility code & Observatory/Mission name & Telescope & Instrument & Pixel scale\\
& & size [m] & & [arcsec/pixel]\\
\hline
\hline
ASAS{-}SN & The All Sky Automated Survey for SuperNovae & $0.14$ & FLI ProLine PL230 & $7.80$ \\
 &\hspace{0.2in}{(global 24-telescope network)} & & & \\
LCO-1m & Las Cumbres Observatory & $1.00$ & Sinistro & $0.39$ \\
 &\hspace{0.2in}{(global 6-telescope network)} & & & \\
\gaia{} & ESA space mission & $1.4\times0.5$ & CCD $4500\times1966$ & $0.20$\\
ZTF & The Zwicky Transient Facility & $1.22$ & CCD $16\times6144\times6160$ & $1.00$\\
ZAO$^{(B)}$ & Znith Astronomy Observatory & $0.20$ & Moravian G2-1600 & $0.99$ \\
Slooh & Slooh (global 10-telescope network) & $0.36, 0.50$ & CCD & $0.63, 0.73$\\
HAO68$^{(B)}$ & Horten Videregaende Skole & $0.68$ & Moravian G2-1600 & $0.79$ \\
AstroLAB{-}IRIS$^{(B)}$ & AstroLAB IRIS & $0.68$ & SBIG STL 6303E & $0.62$ \\
Maidenhead & Commercial telescopes & various & various & various\\
Loiano$^{(B)}$ & Cassini telescope, Loiano Observatory & $1.52$ & BFOSC & $0.58$ \\
Flarestar$^{(B)}$ & Meade SSC-10, Flarestar Observatory & $0.25$ & Moravian G2-1600 & $0.99$ \\
Tacande & Tacande Observatory & $0.40$ & SX814 CCD & $0.29$ \\
pt5m$^{(B)}$ & Roque de los Muchachos Observatory & $0.50$ & QSI-532ws & $0.28$ \\
TJO\_MEIA2$^{(B)}$ & Observatori del Montsec & $0.80$ & MEIA2 CCD & $0.36, 0.36$\\
 ACT{-}452 & 35/51 cm Maksutov telescope, & $0.51, 0.35$ & CCD & $0.36$\\
 & \hspace{0.2in} Mol\.{e}tai Astronomical Observatory & & &\\
GeNAO$^{(B)}$ & SCT-14, Georgian National Astrophysical Observatory & $0.36$ & SXVR-H36 CCD & $0.77$\\
Adonis$^{(B)}$ & Sky-watcher quattro F4 25-cm, Adonis observatory & $0.25$ & Moravian G2 1600 & $1.85$ \\
LCOGT-CTIO-1m$^{(B)}$ & LCOGT\_CTIO100, & $1.00$ & Sinistro\_4K & $0.39$ \\
 &\hspace{0.2in}{Cerro Tololo Inter-American Observatory} & & & \\
MOLETAI-35cm$^{(B)}$ & 35-cm Maksutov telescope, Moletai Observatory & $0.35$ & CCD4710 & $2.20$ \\
OAUJ-CDK500$^{(B)}$ & OAUJ-CDK500, & $0.50$ & Apogee F42 & $0.81$ \\
 &\hspace{0.2in}{\tiny Astronomical Observatory of the Jagiellonian University} & & & \\
RRRT$^{(B)}$ & Fan Mountains Observatory & $0.60$ & SBIG STX16803 CCD & $0.38$ \\
UZPW50$^{(B)}$ & UZPW 50-cm, Entre encinas y estrellas & $0.50$ & Moravian G4-9000 & $0.58$ \\
\hline
\hline
\\
\multicolumn{5}{l}{$(B)$ Telescopes whose data were processed automatically using the Black Hole Target Observation Manager (BHTOM).}\\
\end{tabular}
\\ \vspace{0.1in}
\raggedright

\end{table*}

\section{Event parameters} \label{sec:parameters}
There are several open-source tools to fit microlensing photometry. For \event{}, two tools are used: \pylm{} \citep{Bachelet2017} and \mlm{} \citep{Poleski2019}. Both tools refer to {\it VBBL} \citep{Bozza2010, Bozza2018} during its process of binary lens magnification calculation. We find the best event parameters whose theoretical light curve fits the photometric data using the Monte-Carlo Markov Chain (MCMC) algorithm. The resultant fitted parameters with both tools are fairly agreed with each other. The best fit light curve is shown in Figure~\ref{fig:ssbl}.

\subsection{Model type} \label{subsec:modeltype}
We test both binary-lens and binary-source cases. Under the binary-source assumption, the MCMC chains do not converge. The binary-source assumption fails to fit the anomaly because there are two local maxima and a local minimum during the anomaly. Under the binary-lens assumption, the MCMC chains well converge to the best-fit light curve shown in Figure \ref{fig:ssbl} with reduced $\chi^2\sim1.1-1.6$ depending on the data cleaning extent and parameter variations used in a fit. 

\subsection{Parameter distribution} \label{subsec:eventparams}
The fundamental parameters for the binary lens event are: $t_0$ -- the time of the peak magnitude, $u_0$ -- the impact parameter at time $t_0$, $t_E$ -- the Einstein timescale of the event, $q$ -- the mass ratio of lens components, $s$ -- the separation of lens components, $\alpha$ -- the incident angle of the projected source path with respect to the planetary lens system. Additionally, we consider the angular source radius ($\rho$) for the finite source effect. Parameters $u_0$, $s$, and $\rho$ are relative to the Einstein radius ($\theta_E$).

\begin{figure*}
\centering
\begin{tabular}{cc}
\includegraphics[width=\textwidth/2]{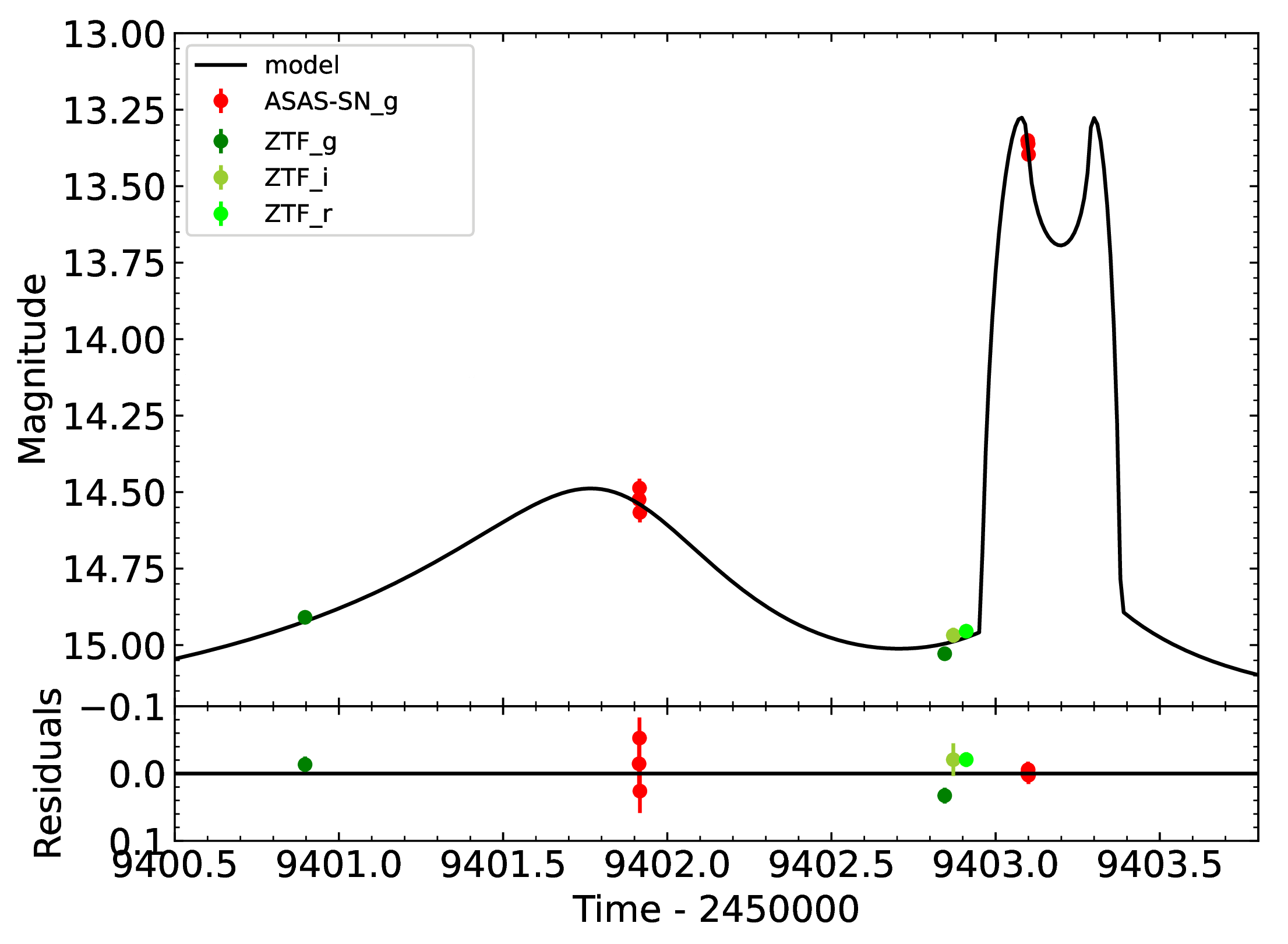} &
\includegraphics[width=\textwidth/2]{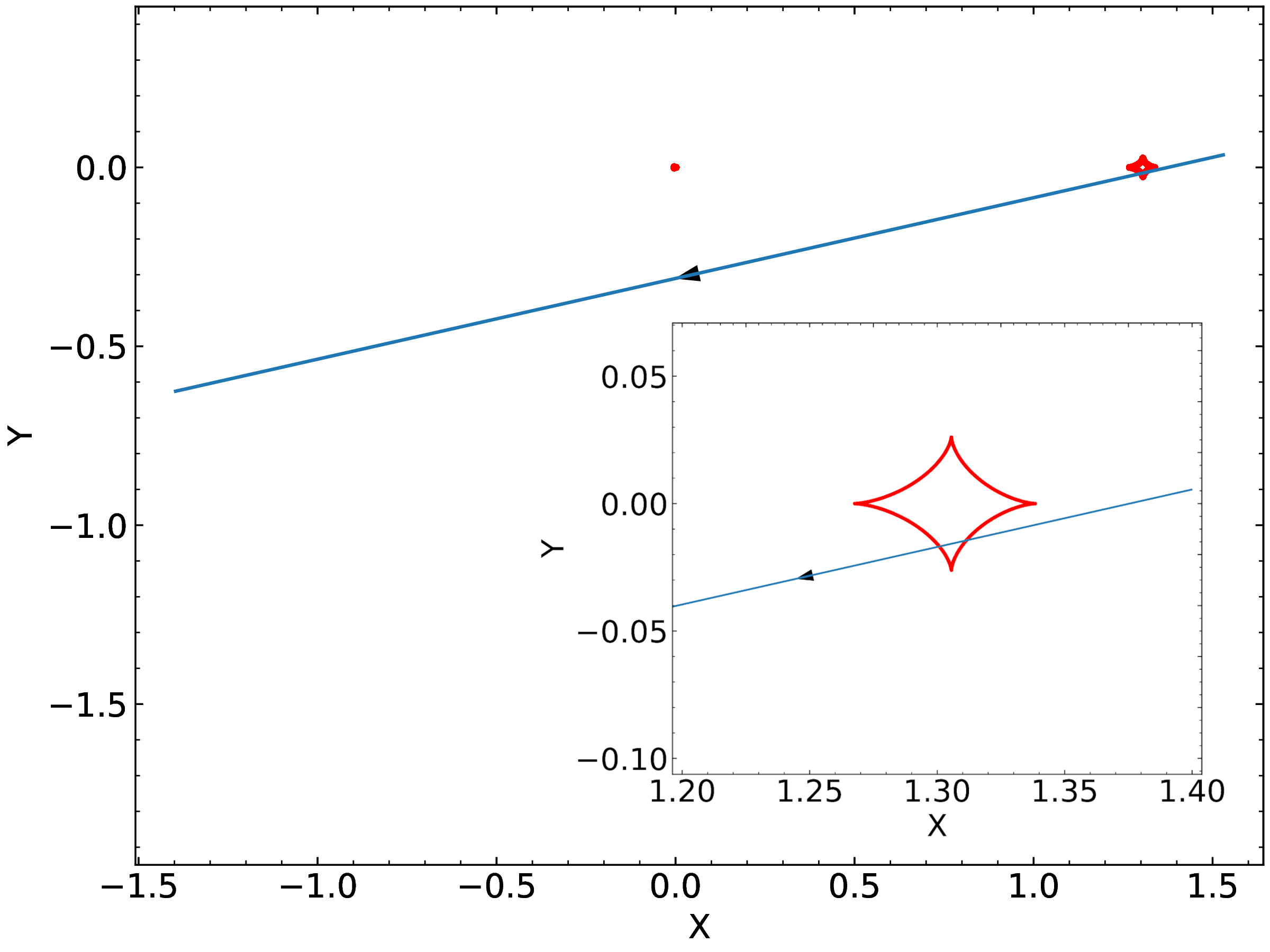} \\
\end{tabular}
\caption{Photometric data and fitted light curve of the anomaly part of the event \event{} (left panel) and the plot of the source trajectory in the lens frame (right panel). Only \asas{} and \ztf{} succeeded to detect the anomaly.}
\label{fig:anomaly}
\end{figure*}

The source trajectory relative to caustics is presented in Figure~\ref{fig:anomaly}. The source first approached the cusp, then crossed the planetary caustic, and passed the central caustic at the end. The model light curve presents characteristic U-shaped signal and the source is fully inside the caustic for a short time. The statistics of event parameter distributions are presented in Table~\ref{tab:params}. All parameter except for the angular source radius ($\rho$) converges to the normal distribution. For $\rho$, there is only an upper limit, which is determined by the just two brightest \asas{} epochs. According to our light curve fitting, the $99\%$ upper limit of $\rho$ is $\sim0.0022$.

We try both wide ($s>1$) and close ($s<1$) models. Close models do not perform very well, and in particular fail to adequately reproduce the high-amplitude \asas{} epochs around HJD2549403.10. 

\renewcommand{\arraystretch}{1.5}
\begin{table}
\centering
\caption{Fitted parameters of the event \event{} (1-sigma uncertainties quoted). For the best fit light curve shown in Figure \ref{fig:ssbl}, the reduced $\chi$ is $\sim$1.6 for 1065 epochs derived by \mlm{} fitting, and $\sim$1.14 for 1220 epochs derived by \pylm{}.}
\label{tab:params}
\begin{tabular}{|l|l|}
	\hline
	Parameters & Values \\ \hline\hline
	$t_0$ (HJD) & $2459438.696\pm0.039$ \\
	$u_0$    & $0.3028\pm0.0025$ \\
	$t_{\mathrm{E}}$ [day] & $27.91\pm0.14$ \\
	$q$      & $(2.611\pm0.088)\times10^{-3}$ \\
	$s$      & $1.8505\pm0.0053$ \\
	$\alpha$ [degree] & $192.700\pm0.076$ \\
	$\rho$   & $(1.25^{+0.47}_{-0.68})\times10^{-3}$ \\
	\hline
\end{tabular}
\end{table}
\renewcommand{\arraystretch}{1}

\subsection{Microlensing parallax and orbital motion} \label{subsec:parallax}

We check if the microlensing parallax ($\vec{\pi_{E}}$) and orbital motion of the lens can be constrained by data. Using \pylm{}, we fit the light curve with three approaches: Static Model (SM) as a basic model of a binary-lens event (described in \S\ref{subsec:eventparams}), Full Parallax Model (FPM) which extends the SM model with the addition of the North and East components of the microlensing parallax, and Full Parallax and Circular Orbital Motion model (FPCOMM) which extends the FPM model with the addition of three parameters to describe the orbital motion of the lens. As a result, the six fundamental parameters of the binary-lens event and $\rho$ result in similar estimates for the SM and FPM models, but in the case of the FPM model, the microlensing parallax parameters are not really constrained (especially $\pi_{EN}$). Concerning the FPCOMM model, we see that errors on all parameters shared with other models are bigger, and there are discrepancies in the values of estimated fundamental parameters with respect to other models. The reduced $\chi^2$ of all models are also similar. The best fit parameters with the SM model shows the reduced $\chi^2 \sim 1.14$ for 1220 epochs whilst both FPM and FPCOMM models only improved the reduced $\chi^2$ as $\sim1.13$ for the same epochs. Thus, we do not detect sufficient evidence of these additional effects and therefore we consider the Static Model for the rest of the study.

\subsection{Source and blending fluxes} \label{subsec:flux}
The blending flux varies between $1.8\%$ and $10.6\%$ of the source flux in the $g$ and $i$ bands, respectively. Table \ref{tab:flux} shows the estimated source and blending magnitude for each filter. Since the line of sight is not toward the galactic bulge, the extinction and the background noise are relatively small. The nearby stars are fairly separated from the source (see \S\ref{sec:observation}) so the blending from the nearby star is negligible. The light curve fitting under a binary source assumption failed, but there is a possibility that the source is an unresolved binary star under the large Renormalised Unit Weight Error (RUWE) value of the \gaia{} observation \citep{Lindegren2021}. Both \gaia{} DR2 and DR3 show large RUWE for the source; hence the large RUWE is likely attributed to the wrong model assumption as a single star and it could be the unresolved binary source system. Even if such a companion exists, it is regarded as being below the unresolved magnitude of the \gaia{} sensitivity (i.e. $\sim$25.7 in $G$-band, \citet{Prusti2016, Brown2018, Brown2021}). Therefore, the blend light derived from the fitting is most likely emitted by the lens system itself. There is an extension of the analysis that the lens system has another bright host beyond the region of the microlensing effect, and the blend light is the sum of the two-host-one-planet system. However, the consideration is excessive since there is no signature so far, and we assume that most blending comes from a one-host-one-planet lens system.

\begin{table}
\centering
\caption{Source and blending magnitudes found through the light curve fitting for different bands.}
\hspace*{-0.25in}
\begin{tabular}{|l|l|l|}
	\hline
	Telescopes and filters & Source magnitude & Blending magnitude \\ \hline\hline
	\asas{}-$g$ & $15.976\pm0.009$ & $20.22\pm0.62$ \\
	\gaia{}-$G$ & $15.543\pm0.010$ & $18.57\pm0.16$ \\
	\lco{}-$g$ (APASS) & $15.969\pm0.014$ & $19.74\pm0.81$ \\
	\lco{}-$i$ (APASS) & $15.282\pm0.012$ & $17.60\pm0.14$ \\
	\lco{}-$i$ (SDSS) & $15.306\pm0.014$ & $17.59\pm0.17$ \\
	\ztf{}-$g$ & $15.960\pm0.011$ & $19.15\pm0.32$ \\
	\ztf{}-$i$ & $15.287\pm0.013$ & $17.51\pm0.19$ \\
	\ztf{}-$r$ & $15.432\pm0.014$ & $18.37\pm0.31$ \\
	\hline
\end{tabular}
\label{tab:flux}
\end{table}

\section{Physical properties of the source and lens} \label{sec:results}

\subsection{Source Properties} \label{subsec:source_analysis}
As Table \ref{tab:source} shows, the RUWE value is not at the acceptable level so that the estimated parallax and distance in Table \ref{tab:source} are uncertain. To find the source properties, we refer to spectroscopic data taken by \ohp{} and \pepsi{} and to the microlensing modelling approach using \pylimass{} \citep{Bachelet2024} and \bes{} Galactic model \citep{Robin2003,Marshall2006,Robin2012,Robin2014}. The results from all approaches are summarised in Table \ref{tab:source_property}.

\subsubsection{\ohp{} data} \label{subsubsec:ohp_analysis}
The \ohp{}\footnote{\url{https://ohp.osupytheas.fr/telescope-de-193cm/}} is a low-resolution spectrograph ($R\sim700$) camera mounted on the 1.93-m telescope at OHP \citep{Schmitt2024}. The target \event{} (labelled as \gaiaevent{} in the \ohp{} data) was observed on 6 September 2021 with an exposure time 900 s. The spectrum was calibrated with ESO-MIDAS\footnote{\url{https://www.eso.org/sci/software/esomidas/}} in a standard way applying bias subtraction, flat-field normalisation as well as wavelength (Hg, Ar, Xe arc lamps) and flux calibration.

\begin{figure}
\centering
\includegraphics[width=0.5\textwidth]{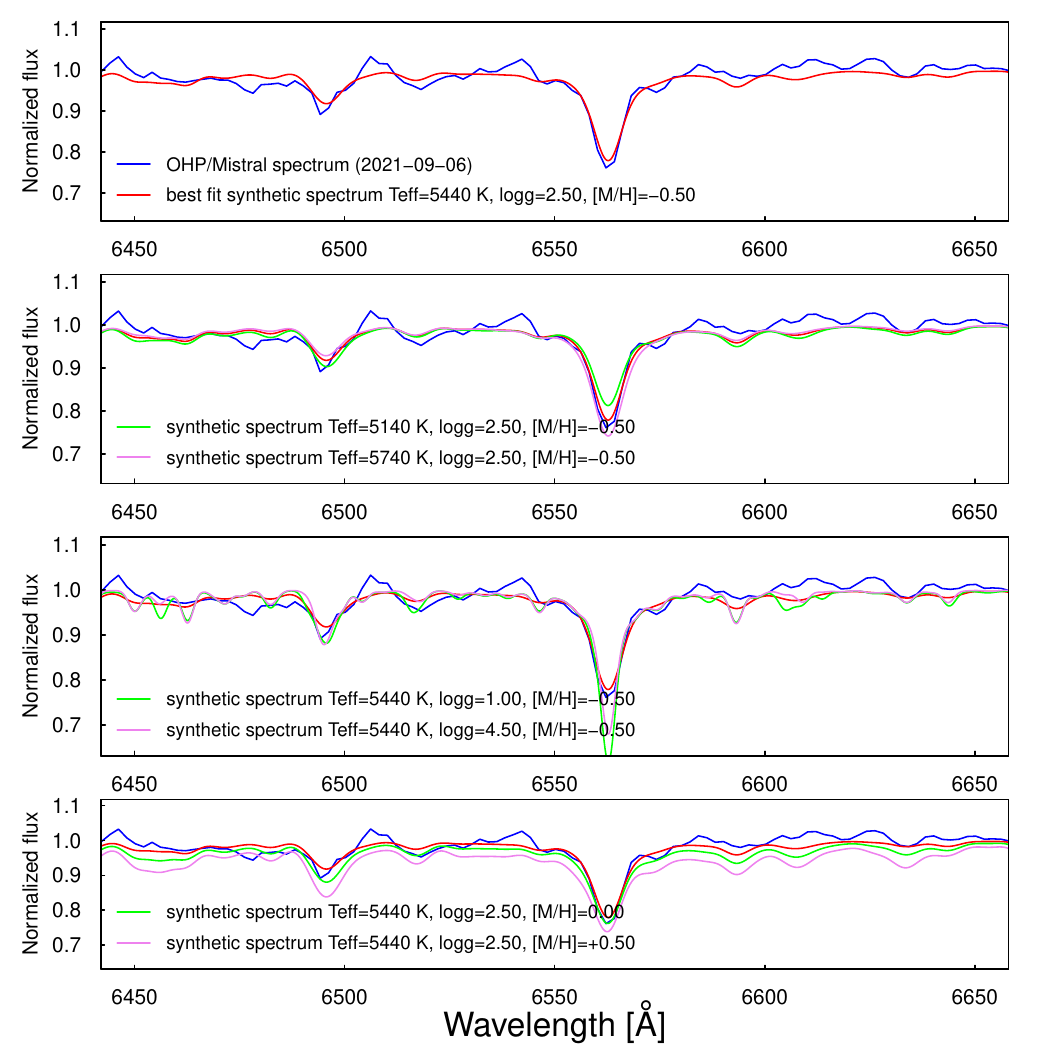} 
\caption{(From the top to bottom) The \ohp{} spectrum (blue) and the best matched synthetic one (red). The same \ohp{} data (blue) but with two other synthetic spectra calculated for different $T_{\rm eff}$ (5140 K -- green, and 5740 K -- pink). The same \ohp{} data (blue) compared with two other synthetic spectra calculated for different $\log g$ (1.00 -- green, 4.50 -- pink).  The same \ohp{} data (blue) compared with two other synthetic spectra calculated for different [M/H] (0.00 -- green, +0.50 -- pink). All plots show the same spectral region around H$\alpha$ line.}
\label{fig:mistral_spectra}
\end{figure}

Using Mistral spectrum, especially H$\alpha$-line region, we analyse the source properties as shown in the second row of Table~\ref{tab:source_property} and Figure~\ref{fig:mistral_spectra}. We fit several synthetic spectra for a range of the parameter set (effective temperature $T_{\rm eff}$, surface gravity $\log g$, metallicity [M/H], and microturbulence velocity $v_{\rm t}$) by using {\it iSpec}\footnote{\url{https://www.blancocuaresma.com/s/iSpec}} framework for spectral analysis \citep{BlancoCuaresma2014,BlancoCuaresma2019} and the SPECTRUM\footnote{\url{http://www.appstate.edu/~grayro/spectrum/spectrum.html}} radiative transfer code. We use the grid of MARCS atmospheric models \citep{Gustafsson2008} and solar abundances provided by \citet{Grevesse2007}. Figure~\ref{fig:mistral_spectra} presents the best fitted synthetic spectrum comparing it with other various theoretical spectra that differ in $T_{\rm eff}$, $\log g$, or [M/H]. It is clearly visible that the models of hotter (5140 K) or cooler (5740 K) star, with lower (1.00) or higher (4.50) surface gravity, or with solar (0.00 dex) or higher than solar (+0.50 dex) metallicity, do not reflect the shape and intensity of the H$\alpha$ spectral region.

Due to the low-resolution of the data, low $SNR\sim65$, and poor weather conditions during observations, we are able to obtain the best fit but with relatively high uncertainties. In general, based on the \ohp{} data, the source star seems to be a metal-poor red giant. Finally, we assume the line-of-sight extinction $A_V=0.21$~mag from \cite{Schlafly2011} and estimate the source distance as $D_s=7.64\pm1.93$~kpc. The resultant source properties are also shown in the 1st row of Table \ref{tab:source_property}.

\subsubsection{\pepsi{} data}	\label{subsubsec:pepsi_analysis}
The \pepsi{}\footnote{\url{https://pepsi.aip.de/}} is a high-resolution ($R>50,000$) optical echelle spectrograph mounted on the Large Binocular Telescope (LBT) with 2x8-m mirrors \citep{Strassmeier2015}. We proposed an observation for the \event{} source using \pepsi{}, and the observation was carried out on 3 June 2023 with an exposure time 3,600 s. \pepsi{} covers the blue arm range with CD-2 configuration (422-479nm) and two red arms range with CD-5 (623-744nm) and CD-6 configuration (735-907nm). This observational material is calibrated using standard PEPSI software for stellar spectroscopy \citep{Ilyin2000}, and therefore, we have obtained high-dispersion spectrum with an S/N ratio of around 70 on the blue part and 174 on the red part.

\begin{figure*}
\centering
\includegraphics[width=0.95\textwidth]{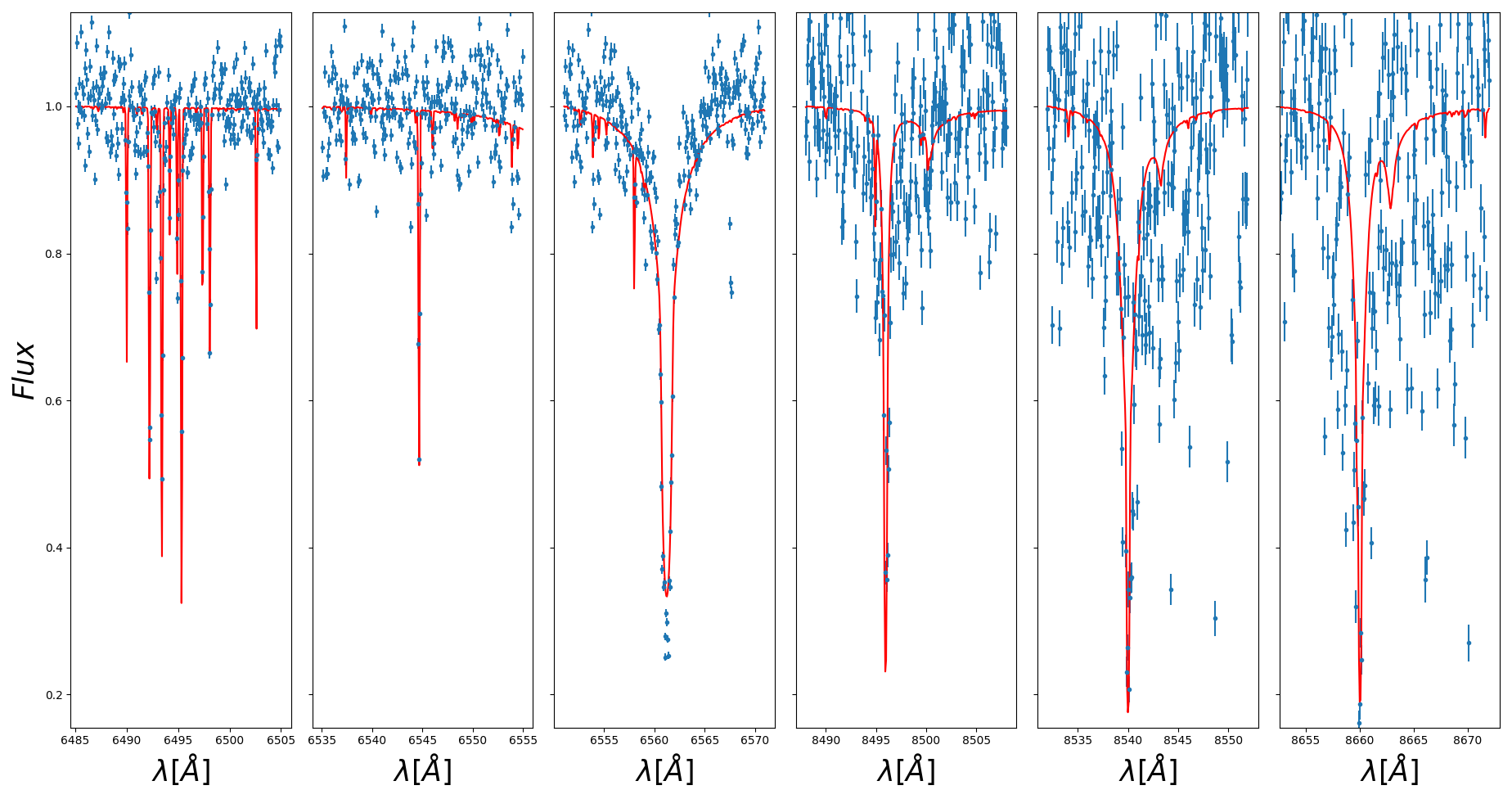} 
\caption{The \pepsi{} spectrum (blue dots) and the best-matched model (red lines). From left to right, the plots show the spectral regions around Ca-I and some other components, Mg-II, H$\alpha$, and three Ca-II lines.}
\label{fig:pepsi_spectra}
\end{figure*}

To fit the \pepsi{} data, we modify the Spyctres algorithm by \citet{Bachelet2022}\footnote{\url{https://github.com/ebachelet/Spyctres/releases}}to support the high-resolution PHOENIX templates \citep{Husser2013} available online \footnote{\url{http://phoenix.astro.physik.uni-goettingen.de/}}. We then model several lines, including the H$\alpha$ line and the calcium triplet lines Ca-II. Some of the spectral regions with the fitted model are shown in Figure \ref{fig:pepsi_spectra}. As summarised in Table~\ref{tab:source_property}, the spectrum lines are best described by a low metallicity red giant model, with $T_{eff}=5330 \pm30$ K, $log(g)=2.22\pm0.07$ cgs, $[Fe/H]=-1.21 \pm0.04$ and a radial velocity $V_r\sim132$ km/s.

Using isochrone models from \mist{} \citep{Dotter2016, Choi2016} with the above parameters and $A_v\sim0.21$ like it is done for \ohp{}, the distance to the source is estimated as $D_S\sim11.4^{+0.59}_{-0.48}$ kpc. The smaller uncertainty than the result of \ohp{} data analysis is attributed to the smaller deviations of the estimated effective temperature, metallicity, and surface gravity.  The relatively metal-rich-weak-gravity estimation of the \pepsi{} data analysis compared with the \ohp{} data analysis results in a farther distance though it is still within the deviation of the \ohp{} result. The resultant source properties are also shown in the 2nd row of Table \ref{tab:source_property}.

\subsubsection{modelling using \pylimass{}}	\label{subsubsec:pylimass_model}
\pylimass{} is an algorithm for finding source-lens properties by analysing the observed data with stellar isochrone models, as described in \citet{Bachelet2024}. In the present case, we use constraints from the lightcurve modelling, namely $t_E$, $\rho$, and the source magnitudes in G, g', and i' bands. We also assume that the blend light measured in the models is emitted by the lens, and therefore we also use the blend light in G, g', and i' bands in the pyLIMASS run. Similarly, we use the 2MASS measurements $J=14.073\pm0.034$ mag and $K_s=13.520\pm0.032$ mag \citep{Cutri2003} as a constraint for the baseline magnitude (considered as the sum of the source and the lens). Finally, we include a constraint on the visual absorption $A_V=0.2\pm0.1$ from the NASA/IPAC Extragalactic Database \footnote{\url{https://ned.ipac.caltech.edu/extinction_calculator}} \citep{Schlafly2011}. The resultant source properties are also shown in the 3rd row of Table \ref{tab:source_property}. The lens distance and mass simultaneously derived via this \pylimass{} modelling are $D_L\sim1.1\pm0.2$ kpc and $M_L\sim0.63\pm0.07$ [$M_{\odot}$].

\subsubsection{modelling using \bes{} Galactic model}	\label{besanson_model}
The \bes{} Galactic model is a simulation tool to generate a stellar catalogue for a given line of sight \citep{Robin2003,Marshall2006,Robin2012,Robin2014}. The stellar catalogue for the simulation is taken from the version m1612\footnote{\url{https://model.obs-besancon.fr}}. The stellar data in the catalogue are generated along the line of sight up to $15.0$ kpc away from the Sun, and the catalogue contains main-sequence stars in MK spectral system (MK), asymptotic giant branch giant stars (AGB), and white dwarfs (WD) with the population ratio of MK:AGB:WD=1568.43:0.00017:0.0177. Using the stellar catalogue, we simulate the microlensing event models focusing on the source properties. 

The \bes{} model uses the Johnson-Cousins filtering system so we consider the magnitude constraints of the source and lens in the $V$ band converted from, the $g$-, $r$-band of \ztf{}. \cite{Tonry2012} and \cite{Medford2020} show the correspondence of the magnitude in $g$- and $r$-bands between \ztf{} and \pan{} and the conversion to some Johnson-Cousins filters. Although the conversion equations of \cite{Tonry2012} are in the Vega system whilst \ztf{}, \pan{} and the \bes{} model are in the AB system, the difference between them in the $V$ band is usually small for the main sequence stars. Using these conversion equations and sample event parameters from the light curve fitting process, we derive the source magnitude $V_S = 15.704 \pm 0.010$, and the blending magnitude $V_b = 19.142 \pm 0.254$. 

We also regard the colour index, proper motion, and a type of star from the \gaia{} data. Using the filter conversion formula \citep{Leeuwen2018} of \gaia{}, we derive the colour in $V-I$ as $0.855 \pm 0.068$ with the correlation between $V$ and $V-I$ as $-1.353\times10^{-3}$. The proper motion in the celestial coordinates is also converted into the galactic coordinates to adjust for the catalogue from the \bes{} model, and we have $(\mu_l,\mu_b) = (-9.160\pm0.035,1.990\pm0.041)$ with the correlation of $-0.419$. The constraint of the proper motion is not directly applied but applied with the distribution of possible lens proper motions in forms of a magnitude of the relative proper motions. The \gaia{} data also indicates that the source is likely to be a red giant or a sub-giant because the colour is relatively redder with high surface gravity, but a hotter effective temperature than the typical Main Sequence $K-$stars, $M-$dwarfs or so. These values from \gaia{} are based on the low resolution of BP/RP spectra and are suspicious like a parallax value, but the hotter tendencies are plausible. Therefore, we apply the initial cut-off for the effective temperature range (4,250 < $T_{eff}$ < 6,000 [K]) and an upper cut-off for the surface gravity (log(g) < 4.0) \citep{Hekker2011}, and the cut-offs are being adjusted during the effective sampling process using \bes{} catalogues.

By applying the above constraints and the event likelihood from the light curve fitting (Table \ref{tab:params} and Table \ref{tab:flux}) to the \bes{} catalogue, we model the source stars and confirmed that both a red giant and a sub-giant are possible as we expected. The most probable case is the metal-poor red giant according to the weights of tested samples, and its properties fairly agree with the results from the other three approaches (\ohp{}, \pepsi{}, and \pylimass{}). In this modelling, the red giant source tends to be located at $Ds\sim 13.6^{+2.89}_{-3.86}$ kpc. The resultant source properties are also shown in the 4th row of Table \ref{tab:source_property}.

\subsubsection{Summary of the source properties}
From all approaches to analysing the source properties, the metal-poor red giant is the absolute answer for the stellar type. The estimated distance is also fairly agreeable among the approaches as overlapping and/or scratching the deviations. Because the line of sight for \event{} towards out of the bulge, such a farther distance indicates that the source is in the halo with high probability. The halo stars are generally old, and this point also supports the estimated low metallicity. Thus, these four approaches are acceptable, and we determine the final result of the source properties by combining these estimations as shown in the bottom row of Table \ref{tab:source_property}. We assume the results from \pepsi{} data are the most reliable and realistic solution, and we weight the results from the other approaches by relative Q-functions to the \pepsi{}'s deviation. The values from the event models (\pylimass{} and \bes{}) are treated more for reference purposes during finalising the solution since the direct observation of spectra is much more reliable for the stellar parameter estimation. With the final solution in Table \ref{tab:source_property} and the stellar catalogue from the \bes{} model, we find the radius of the source is $7.58^{+2.29}_{-1.83}R_{\odot}$ with the absolute $V$-magnitude of $0.13^{+1.14}_{-0.61}$.

\renewcommand{\arraystretch}{2}
\begin{table*}
\centering
\caption{Source properties found from spectroscopic data and the Galactic model.}
\begin{tabular}{|l|cccc|}
	\hline
	Spectrum / Model & $T_{eff} [K]$ & [Fe/H] & $\log (g)$ [cgs] & $D_S (kpc)$ \\ \hline\hline
	\ohp{} & $5440\pm300$ & $-0.77\pm0.30$ & $2.50\pm0.50$ & $7.64\pm1.93$ \\ \hline
	\pepsi{} & $5330\pm30$ & $-1.21\pm0.04$ & $2.22\pm0.07$ & $11.4^{+0.59}_{-0.48}$ \\ \hline\hline
	\pylimass{} & $5500\pm140$ & $-1.50\pm0.40$ & $2.60\pm0.10$ & $11.0\pm1.0$ \\ \hline
	\bes{} & $4881^{+144}_{-97}$ & $-1.57^{+0.46}_{-0.51}$ & $2.20^{+0.37}_{-0.21}$ & $13.6^{+2.89}_{-3.86}$ \\ \hline\hline
	Final solution & $5384^{+135}_{-128}$ & $-1.21^{+0.05}_{-0.05}$ & $2.28^{+0.30}_{0.31}$ & $11.8^{+0.76}_{-0.49}$ \\ \hline
\end{tabular}\\
\label{tab:source_property}
\vspace{6pt}
Note: The final solution at the bottom row is the combined result of the above four approaches.
\end{table*}
\renewcommand{\arraystretch}{1}

\subsection{Lens results} \label{subsec:lens_analysis}
Using the \bes{} catalogue, we simulate the event once again by focusing on the lens properties. The event parameters from \S\ref{sec:parameters} and the source properties from \S\ref{subsec:source_analysis} are applied as the constraints during the event sampling. The luminosity, radius, and effective temperature of a lens host star are calculated using the mass relationships presented by \cite{Cuntz2018} and \cite{Parsons2018}. For the lens-planet system, we assume that the eccentricity and the longitude of the ascending node of the lens system are zero. Since the projected separation between the lens host and planet is larger than the Einstein radius, we can assume that the physical separation is also large enough to make the orbital motion effect negligible with respect to the Einstein timescale. In this case, any input of the projected position of the planet in the sky can be covered by the inclination of the orbital plane and the current position (i.e. true anomaly), and we can simplify the eccentricity and the longitude of the ascending node. Then, the semi-major axis is derived from the inclination, true anomaly, distance to the system ($D_L$), and the projected separation of the host and planetary lenses ($s$) in the observer frame. Each set of event parameters is tested 10 times by randomly selecting the inclination and true anomaly every time to reduce sampling noise.

\begin{figure}
\centering
\vspace{-0.2in}
\includegraphics[height=0.3\textheight]{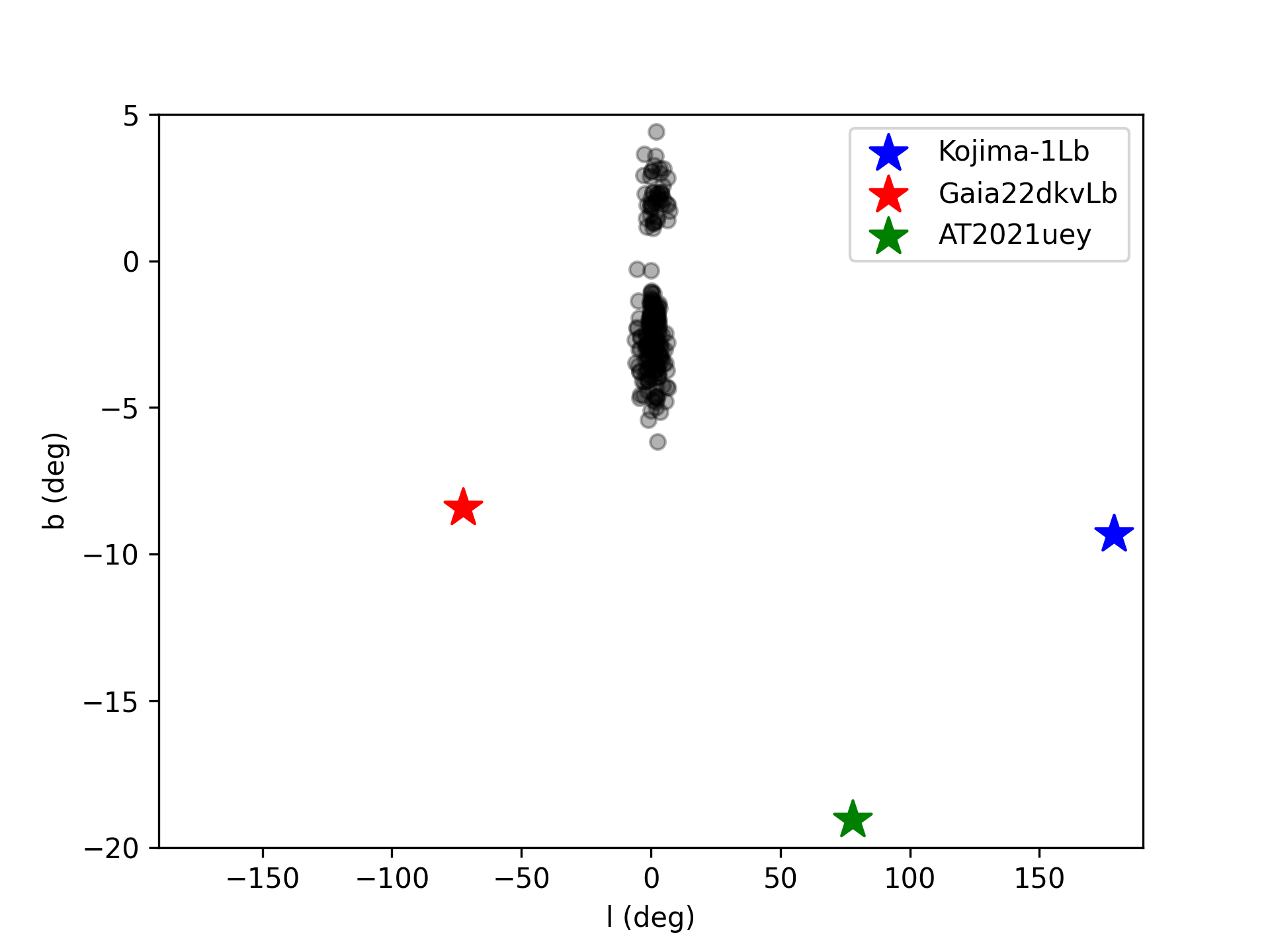}\\
\includegraphics[height=0.3\textheight]{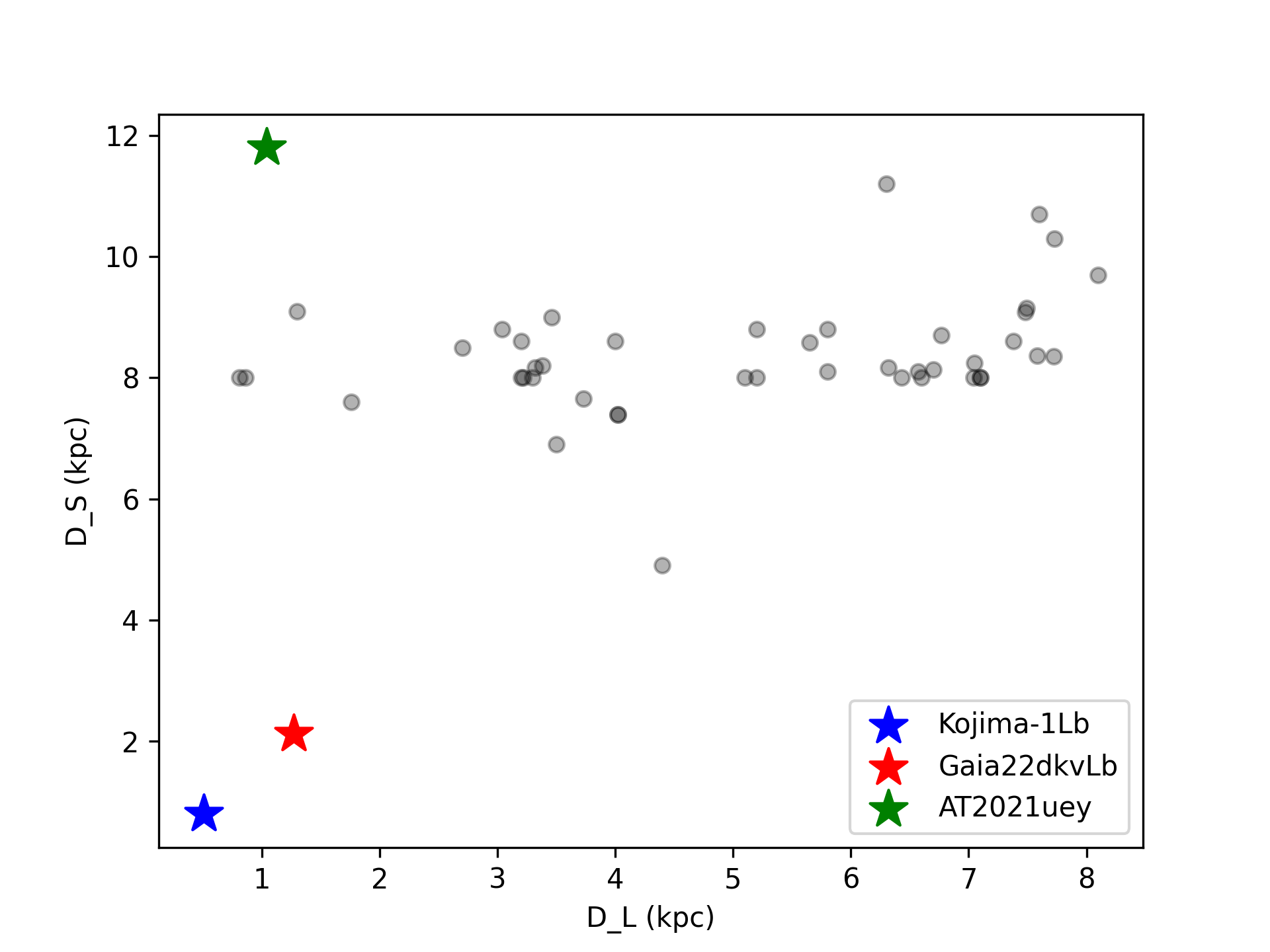}\\
\includegraphics[height=0.3\textheight]{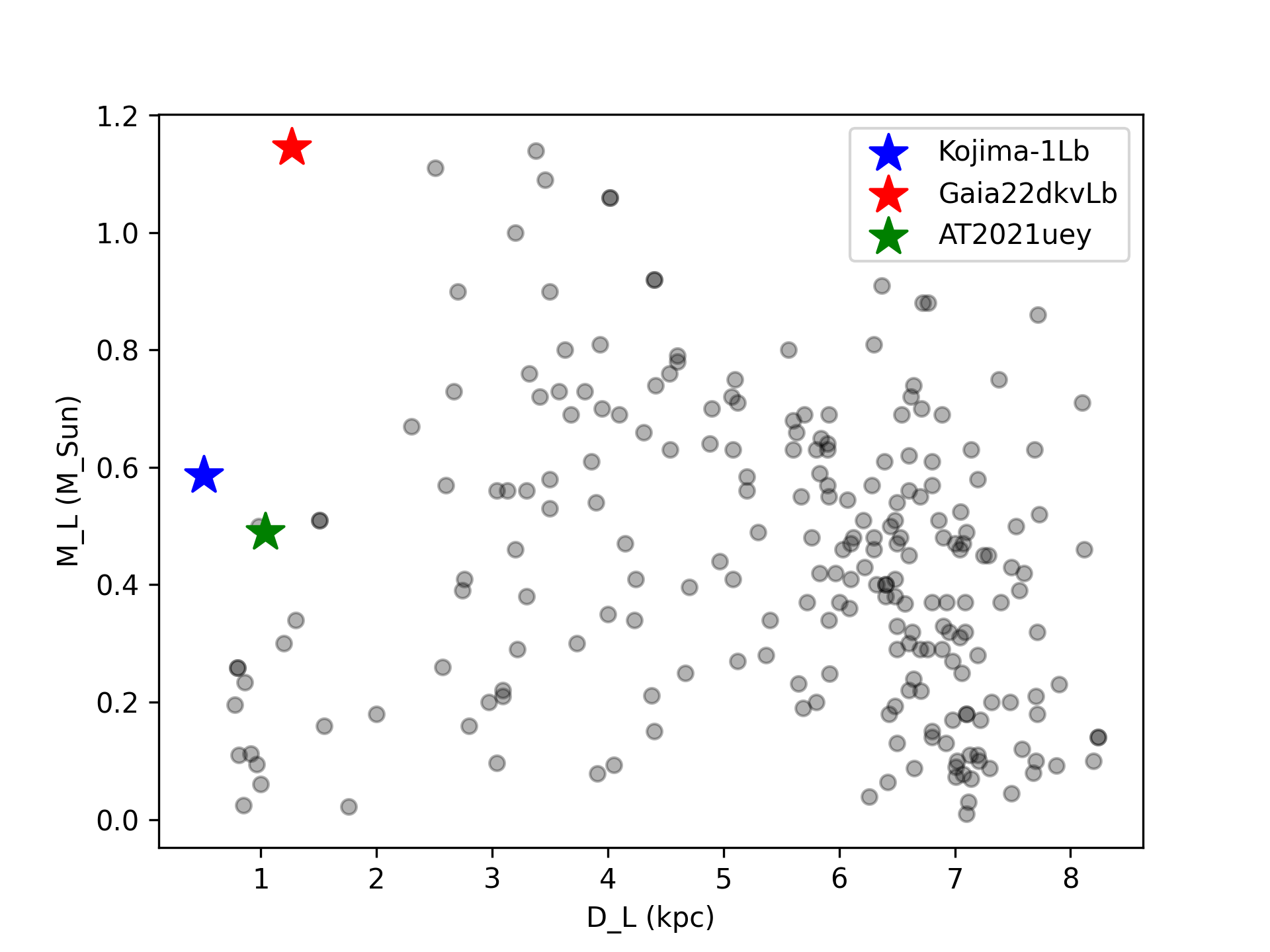}\\
\caption{Comparison for the exoplanet microlensing event distribution in the Galactic coordinates (top), the lens-source distance relation (middle), and the lens distance-mass relation (bottom). The coloured markers are the events outside the bulge as labelled in each panel, and the dark dots are the exoplanet microlensing events toward the bulge. Data is taken from the NASA Exoplanet Archive. Note that the scale on the top panel is very different between the longitude ($l$) and latitude ($b$) axes.}
\label{fig:distribution}
\end{figure}

Table \ref{tab:lens_sampling} summarises the weighted mean and deviations for the lens parameters. Since our observation target is not toward the Galactic bulge, it is plausible that the lens system is likely located close to us. The \pylimass{} modelling discussed in \S\ref{subsubsec:pylimass_model} results in almost the same distance to the lens. About a half solar mass is also roughly agreed between the results from the \bes{} and \pylimass{} approaches. Such a mass and the low temperature indicate that the lens star is an M-dwarf. The snowline of such dwarf star is generally $< 1$ AU \citep{Mulders2015}, therefore the derived separation of the lens system indicates that the lens star has a Jupiter-like companion beyond its snowline. With the estimated lens distance and mass, the spatial visualisation of \event{} compared with Kojima-1, \dkv{}, and other exoplanet microlensing events toward the bulge are summarised in Figure \ref{fig:distribution}.

\renewcommand{\arraystretch}{2}
\begin{table}
\centering
\caption{Summary of the lens properties derived from the event simulation.}
\begin{tabular}{|l|l|}
    \hline
	Property & Value \\ \hline\hline
	$D_L [kpc]$ & $1.04^{+0.74}_{-0.44}$\\
	$M_* [M_{\odot{}}]$ & $0.49^{+0.16}_{-0.18}$\\
	$M_{V,L}$ & $9.88^{+1.39}_{-1.33}$\\
	$T_{eff,L} [K]$ & $3680^{+307}_{-204}$\\
	$M_{pl} [M_{Jup}]$ & $1.34^{+0.45}_{-0.50}$\\
	$a [AU]$ & $4.01^{+1.68}_{-1.34}$\\
	$log_{10}(P) [day]$ & $3.62^{+0.23}_{-0.23}$\\ \hline
\end{tabular}
\label{tab:lens_sampling}
\end{table}
\renewcommand{\arraystretch}{1}

\section{Conclusions} \label{sec:conclutions}
We estimate that the lens of the event \event{} is an M-dwarf ($M_*\sim0.49^{+0.16}_{-0.18}$ [$M_{\odot}$]) with a Jupiter-like planet ($M_{pl}\sim1.34^{+0.45}_{-0.50}$ [$M_{Jup}$]) beyond the snowline. The derived semi-major axis and the orbital period are the optimised values by ignoring the eccentricity and orbital phase. The planet may move within the snow line if it has a strict eccentricity, but it is still true that the planet spends the most period of time under a frozen circumstance. 
Since the combination of ASAS-SN and ZTF data successfully shows a unique feature of the anomaly (Figure \ref{fig:anomaly}), we have been able to fit the light curve properly and confirm the existence of a planet beyond the snowline. On the other hand, there have been some issues identifying the source properties. By applying several stellar observations and modelling, we have finalised that the source star is likely located in the halo. Since the other binary-lens event examples toward out of the Galactic bulge (\kojima{} and \dkv{}) are disc-disc events, \event{} is the third but a unique example as a disc-halo event that was successfully observed.

\begin{acknowledgements}
Work by Makiko Ban and Rados\l{}aw Poleski was supported by Polish National Agency for Academic Exchange grant ``Polish Returns 2019.''
This work was developed within the Fink community and made use of the Fink community broker resources. Julien Peloton thanks LSST-France and CNRS/IN2P3 for supporting Fink.
We acknowledge ESA \gaia{}, DPAC and the Photometric Science Alerts Team\footnote{\url{http://gsaweb.ast.cam.ac.uk/alerts}}.
Josep Manel Carrasco was (partially) supported by the Spanish MICIN/AEI/10.13039/501100011033 and by "ERDF A way of making Europe" by the “European Union” through grant PID2021-122842OB-C21, and the Institute of Cosmos Sciences University of Barcelona (ICCUB, Unidad de Excelencia ’Mar\'{\i}a de Maeztu’) through grant CEX2019-000918-M. The Joan Oró Telescope (TJO) of the Montsec Observatory (OdM) is owned by the Catalan Government and operated by the Institute for Space Studies of Catalonia (IEEC).
Erika Pak\u{s}tien\.{e}, Justas Zdanavi\u{c}ius, Marius Maskoli\={u}nas, Vytautas \u{C}epas, R\={u}ta Urbonavi\u{c}i\={u}t\.{e}, Edita Stonkut\.{e} acknowledge the support of the observations at Mol\.{e}tai AO from the Research Council of Lithuania (grant No. S-LL-24-1).
Work by Akihiko Fukui and Norio Narita was partly supported by JSPS KAKENHI Grant Number JP17H02871, JP17H04574, JP18H05439, and JST CREST Grant Number JPMJCR1761.
Yiannis Tsapras acknowledges the support of DFG priority program SPP 1992 “Exploring the Diversity of Extrasolar Planets” (TS 356/3-1).
The project is supported by the European Union's Horizon 2020 research and innovation program under grant agreement 101004719.
We are grateful to Kornel Howil and Maja Jab\l{}o\'{n}ska who created a simulation code for a low blending microlensing event (so-called {\it Dark Lens Code}, \citet{Howil2025}) and offered the code on our demand though we have not used it in final analysis. 
We acknowledge the support from the European Union's research and innovation programmes under grant agreements No 101004719 (OPTICON-RadioNet Pilot, ORP) and No 101131928 (ACME). We acknowledge the support from the Polish National Science Centre NCN grant DAINA No 2024/52/L/ST9/00210.
This work was partially supported by a program of the Polish Ministry of Science under the title ‘Regional Excellence Initiative’, project no. RID/SP/0050/2024/1.
Based on observations made at Observatoire de Haute Provence (CNRS), France, with MISTRAL on the T193 telescope. The LBT is an international collaboration among institutions in the United States, Italy and Germany. LBT Corporation partners are The Uni- versity of Arizona on behalf of the Arizona university system; Istituto Nazionale di Astrofisica, Italy; LBT Beteiligungsgesellschaft, Germany, representing the Max- Planck Society, the Astrophysical Institute Potsdam, and Heidelberg University; The Ohio State University, and The Research Corporation, on behalf of The Uni- versity of Notre Dame, University of Minnesota and the University of Virginia. This project used data obtained via BHTOM (https://bhtom.space), which has received funding from the European Union’s Horizon 2020 research and innovation programme under grant agreements No. 730890 (OPTICON) and 101004719 (ORP) and Horizon Europe programme No. 101131928 (ACME).
R.F.J. acknowledges support for this project provided by ANID's Millennium Science Initiative through grant ICN12\textunderscore 009, awarded to the Millennium Institute of Astrophysics (MAS), and by ANID's Basal project FB210003
\end{acknowledgements}

\bibliographystyle{aa}
\bibliography{ref}

\end{document}